\documentclass[epj,twocolumn]{webofc}
\usepackage[varg]{txfonts} 
\usepackage{amsmath,amssymb}
\usepackage{slashed}
\usepackage{widetext}
\allowdisplaybreaks
\newcommand{\be}{\begin{eqnarray}}
\newcommand{\ee}{\end{eqnarray}}
\usepackage{hyperref}
\usepackage{cancel}
\newcommand{\nn}{\nonumber}

\woctitle{Transversity 2014}
\begin{document}
\title{Wigner Functions and Quark Orbital Angular Momentum }
\author{
Asmita Mukherjee\inst{1}\fnsep\thanks{\email{asmita@phy.iitb.ac.in}} \and
Sreeraj Nair\inst{1}\fnsep\thanks{\email{sreeraj.phy@gmail.com}} \and
Vikash Kumar Ojha\inst{1}\fnsep\thanks{\email{ohjavikash@gmail.com}}
}
\institute{Department of Physics, Indian Institute of Technology Bombay, Powai, Mumbai 400076, India
}
\abstract{Wigner distributions contain combined position and momentum space information of
the quark
distributions and are related to both generalized parton distributions 
(GPDs)  and transverse momentum dependent parton distributions (TMDs). 
We report on a recent model calculation of the Wigner distributions for the
quark and their relation to the orbital angular momentum.}  
\maketitle
\section{Introduction}

\noindent
A fundamental question in hadronic physics is to understand how the spin of
the nucleon is divided among the quarks and gluons that form the nucleon. In
the EMC experiment, it was found that only a small fraction of the nucleon
spin is carried by the quarks and antiquarks. Recent experiments suggest that 
the intrinsic spin carried by the gluons is also small. Thus a substantial
part of the spin comes from quark and gluon orbital angular momentum (OAM). There
are issues related to the gauge invariance and experimental measurability  
that complicates the understanding of the OAM. However, recently it was
shown that the quantum mechanical Wigner distributions of quarks inside the
nucleon can give information on the OAM carried by the quarks. Wigner
distributions can be thought of as so-called quantum mechanical phase space
distributions which give a joint position and momentum space information
about the quarks in the nucleon. As position and momentum operators do not
commute in quantum mechanics, they cannot be simultaneously determined. As a
result, Wigner distributions are not positive definite. However, a reduced
Wigner distribution can be defined after integrating over several variables,
and these  are positive definite. The Wigner distributions are related to the
generalized parton correlation functions (GPCFs) or generalized transverse
momentum dependent parton distributions (GTMDs). These are the mother
distributions of the GPDs and TMDs, both of which contain very useful
information on the 3D structure of the nucleon as well as the spin and OAM
of the quarks in the nucleon. In \cite{lorce} the authors introduced 5-D reduced
Wigner distributions in the infinite momentum frame, or in light-front
framework, which are functions of
three momentum and two position variables. Working in the light-front
formalism is useful as the transverse boosts are Galilean or do not involve
dynamics and longitudinal boosts are scale transformations.  Thus it is
possible to have a boost invariant description of the Wigner distributions
in terms of light-front wave functions. In \cite{lorce} the Wigner distributions were
calculated in light-cone constituent quark model and light-cone chiral
quark-soliton model. Both these models have no gluonic degrees of freedom. 
In this work \cite{us}, we calculate the Wigner functions for a dressed quark in
light-front Hamiltonian perturbation theory, which is basically a
relativistic composite spin $1/2$ state. This is a simple model with a
gluonic degree of freedom. The state in expanded in Fock space in terms of
multiparton light-front wave functions (LFWFs). The  advantage is that  
this approach gives a field theoretic description of deep inelastic
scattering processes and at the same time keeps close contact with parton
model, the partons now are field theoretic, they are massive, non-collinear
and interacting. To obtain the non-perturbative LFWFs for a bound state like
the nucleon one needs a model light-front Hamiltonian. However, for a quark
state dressed with a gluon the two-body light-front wave function can be
obtained analytically. In the next section, we present a calculation of the
Wigner distributions in this model.     
\section{Wigner Distributions}

\noindent
The Wigner distribution of quarks can be defined as
\cite{lorce,metz}

\be \label{main}
\rho^{[\Gamma]} ({b}_{\perp},{k}_{\perp},x,\sigma) = \int \frac{d^2
\Delta_{\perp}}
{(2\pi)^2} e^{-i \Delta_{\perp}.b_{\perp}}\nn \\
W^{[\Gamma]} (\Delta_{\perp},{k}_{\perp},x,\sigma);
\ee

\noindent
where  $\Delta_{\perp}$ is momentum transfer from the initial state to the
final state in transverse
direction and
${b}_{\perp}$ is 2 dimensional vector
 in impact parameter space conjugate to $\Delta_{\perp}$. $W^{[\Gamma]}$ is
the quark-quark correlator given by

\begin{flalign} \label{qqc}
W^{[\Gamma]} ({\Delta}_{\perp},{k}_{\perp},x,\sigma)  =
\frac{1}{2}\int\frac{dz^{-}d^{2} z_{\perp}}{(2\pi)^3}\nonumber\\e^{i(xp^+
z^-/2-k_{\perp}.z_{\perp})}\nonumber\\
 \Big{\langle } p^{+},\frac{\Delta_{\perp}}{2},\sigma \Big{|}
\overline{\psi}(-\frac{z}{2}) \Omega\Gamma \psi(\frac{z}{2}) \Big{|}
p^{+},-\frac{\Delta_{\perp}}{2},\sigma \Big{\rangle }
\Big{|}_{z^{+}=0}.
\end{flalign}

\noindent
We take the target state to be a quark dressed with a gluon. We use the symmetric frame
\cite{brodsky}
where $p^+$ and $\sigma$ define the longitudinal momentum of the target state and its
helicity respectively. $x = k^+/p^+$ is the fraction of
longitudinal momentum of the dressed quark carried by the quark. 
$\Omega$ is the gauge link needed for color gauge invariance. Here, we use the
light-front gauge, $A^+=0$  and take the gauge link to be unity. In fact the
quark orbital angular momentum in this model does not depend on the gauge
link. The symbol $\Gamma$
represents the Dirac matrix structure.
The state of momentum $p$ and helicity $\sigma$,
can be expanded in Fock space using the 
multi-parton light-front wave functions (LFWFs) \cite{hari}. The boost
invariant two-particle LFWFs  be calculated perturbatively as \cite{hari}.
We use the two component formalism \cite{zhang}. 
The quark state dressed with  a gluon as we consider here mimics the bound state of 
a spin-1/2 particle and a spin-1 particle. However, for such a bound state the bound state mass $M$
should be less than the sum of the masses of the constituents for stability.
In this work, we use the same mass for the bare
as well as the dressed quark in perturbation theory \cite{hari}. The
Wigner distributions can be expressed as overlaps of two-particle LFWFs. We take the
momentum transfer to be completely in the transverse direction. In this
case, the overlaps are diagonal or between two-particle LFWFs.\\

\noindent
Wigner distributions of quarks with longitudinal polarization $\lambda$ in a
target with longitudinal polarization $\Lambda$ is denoted  by $
\rho_{\Lambda \lambda}(\vec{b}_\perp,\vec{k}_\perp,x)$. This can be
decomposed in terms of four Wigner functions as defined below: 

\be \label{ruu}
\rho_{UU}(\vec{b}_\perp,\vec{k}_\perp,x) = \frac{1}{2}\Big[\rho^{[\gamma^+]}
(\vec{b}_\perp,\vec{k}_\perp,x,+\vec{e}_z) \nonumber\\ +
\rho^{[\gamma^+]}(\vec{b}_\perp,\vec{k}_\perp,x,-\vec{e}_z) \Big]
\ee
is the Wigner distribution of unpolarized quarks in unpolarized target
state.

\be \label{rlu}
\rho_{LU}(\vec{b}_\perp,\vec{k}_\perp,x)= \frac{1}{2}\Big[\rho^{[\gamma^+]}
(\vec{b}_\perp,\vec{k}_\perp,x,+\vec{e}_z) \nonumber\\ -
\rho^{[\gamma^+]}(\vec{b}_\perp,\vec{k}_\perp,x,-\vec{e}_z) \Big]
\ee
is the distortion due to longitudinal polarization of  the target state.

\be \label{rul}
\rho_{UL}(\vec{b}_\perp,\vec{k}_\perp,x) =
\frac{1}{2}\Big[\rho^{[\gamma^+\gamma_5]}
(\vec{b}_\perp,\vec{k}_\perp,x,+\vec{e}_z)\nonumber\\ +
\rho^{[\gamma^+\gamma_5]}(\vec{b}_\perp,\vec{k}_\perp,x,-\vec{e}_z) \Big]
\ee
is the distortion due to the longitudinal polarization of quarks, and

\be \label{rll}
\rho_{LL}(\vec{b}_\perp,\vec{k}_\perp,x) =
\frac{1}{2}\Big[\rho^{[\gamma^+\gamma_5]}
(\vec{b}_\perp,\vec{k}_\perp,x,+\vec{e}_z)\nonumber\\-
\rho^{[\gamma^+\gamma_5]}(\vec{b}_\perp,\vec{k}_\perp,x,-\vec{e}_z) \Big]
\ee
is the distortion due to the correlation between the longitudinal
polarized target state and quarks. $+\vec{e_z}$ and $-\vec{e_z}$ denote the 
helicity up and down of the target state, respectively.  
In our model, $\rho_{LU} = \rho_{UL}$.\\

\begin{figure}[!htp]
\centering
\includegraphics[width=6cm,height=6cm,clip]{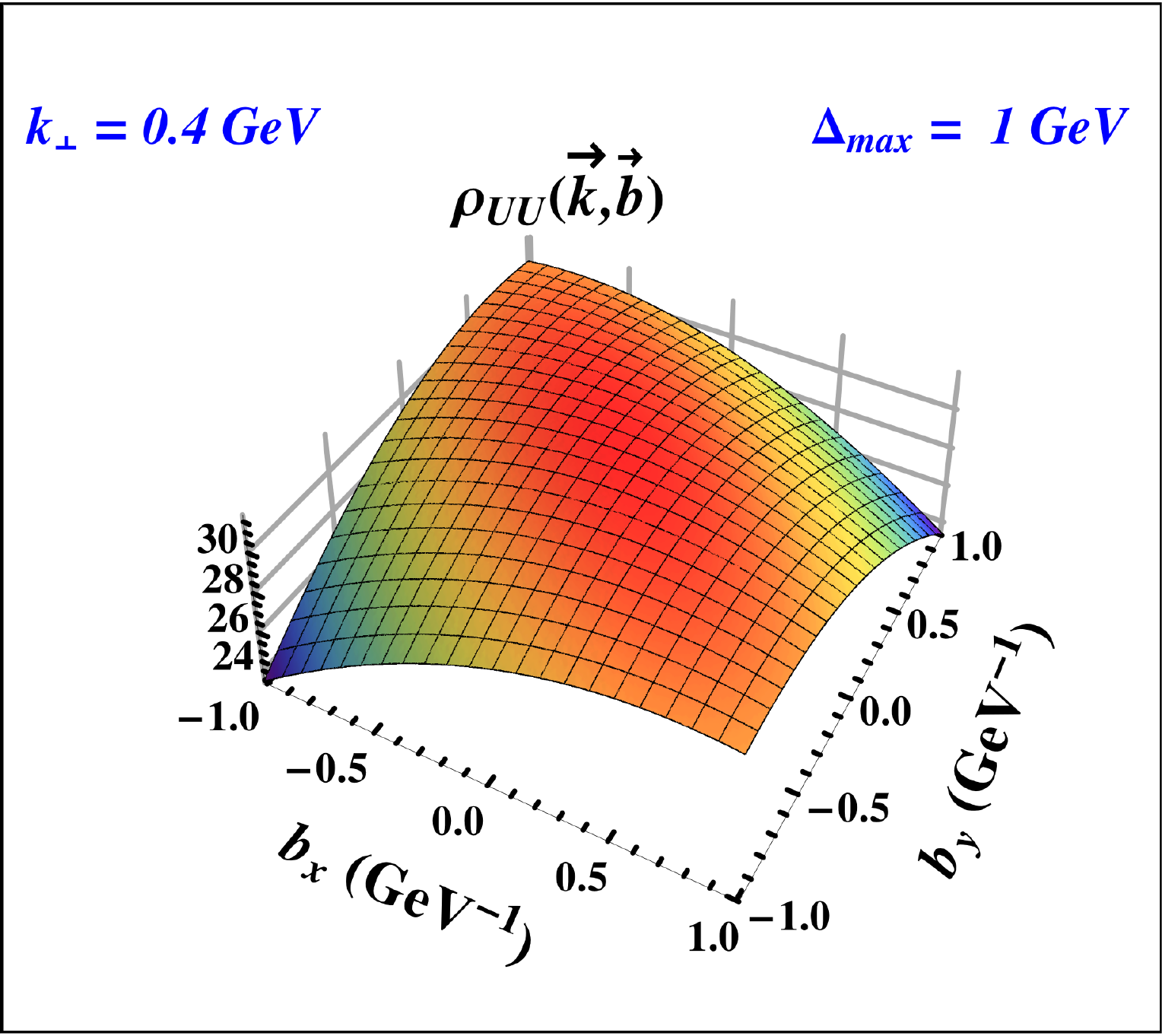}
\caption{\label{fig1}(Color online)
3D plots of the Wigner distributions $\rho_{UU}$  in
$b$ space for $\Delta_\perp^{max} = 1.0$ GeV with $k_\perp = 0.4$ GeV.
For all the plots we kept $m = 0.33$ GeV, integrated out the $x$ variable
and we took $\vec{k_\perp} 
= k \hat{j}$ and $\vec{b_\perp} = b \hat{j}$.  }
\end{figure}

\begin{figure}[!htp]
\centering
\includegraphics[width=6cm,height=6cm,clip]{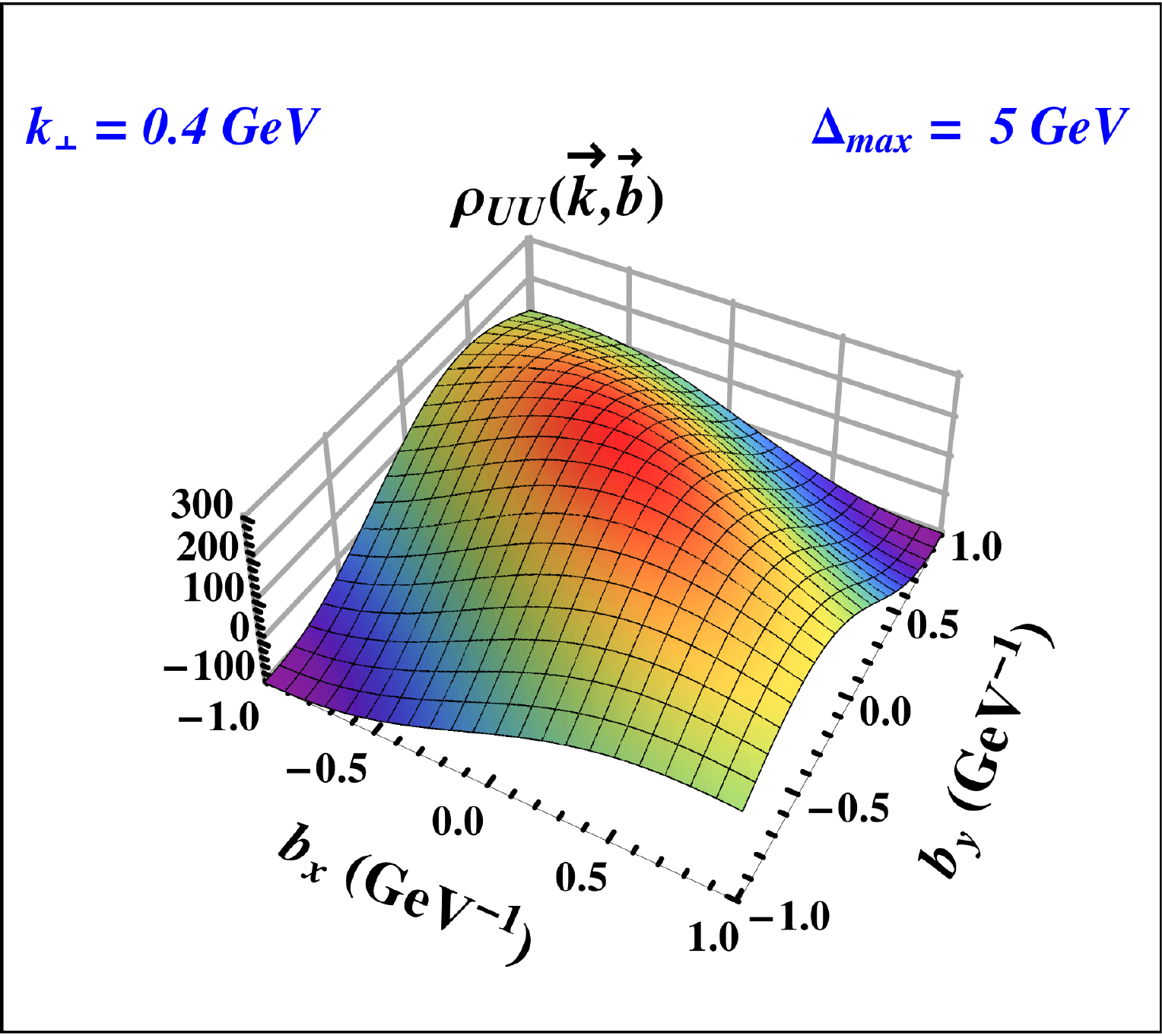}
\caption{\label{fig2}(Color online)
3D plots of the Wigner distributions $\rho_{UU}$ in
$b$ space for $\Delta_\perp^{max} = 5.0$ GeV with $k_\perp = 0.4$ GeV.}
\end{figure}

\begin{figure}[!htp]
\centering
\includegraphics[width=6cm,height=6cm,clip]{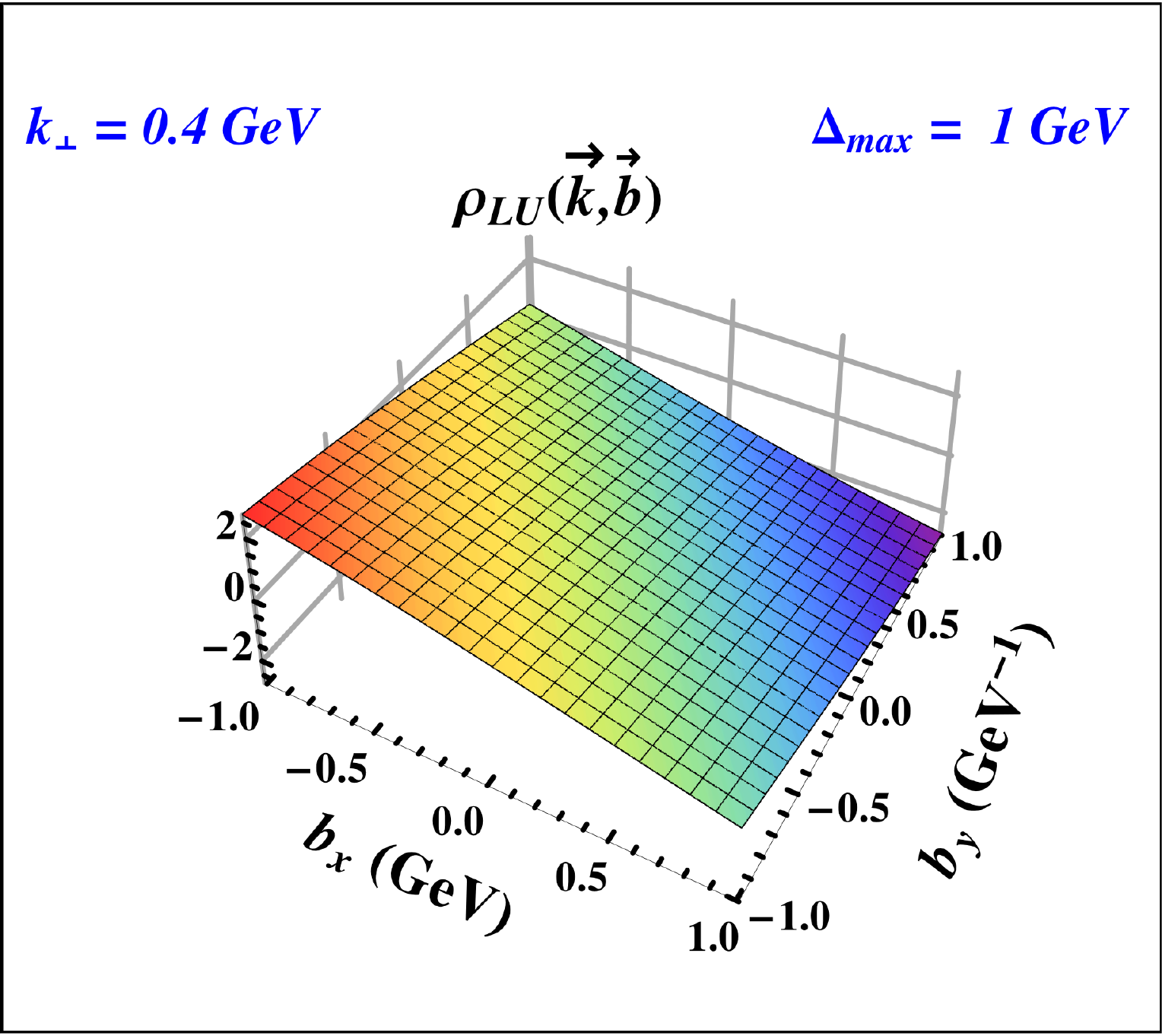}
\caption{\label{fig3}(Color online)
3D plots of the Wigner distributions $\rho_{LU}$  in
$b$ space for $\Delta_\perp^{max} = 1.0$ GeV with $k_\perp = 0.4$ GeV. }
\end{figure}

\begin{figure}[!htp]
\centering
\includegraphics[width=6cm,height=6cm,clip]{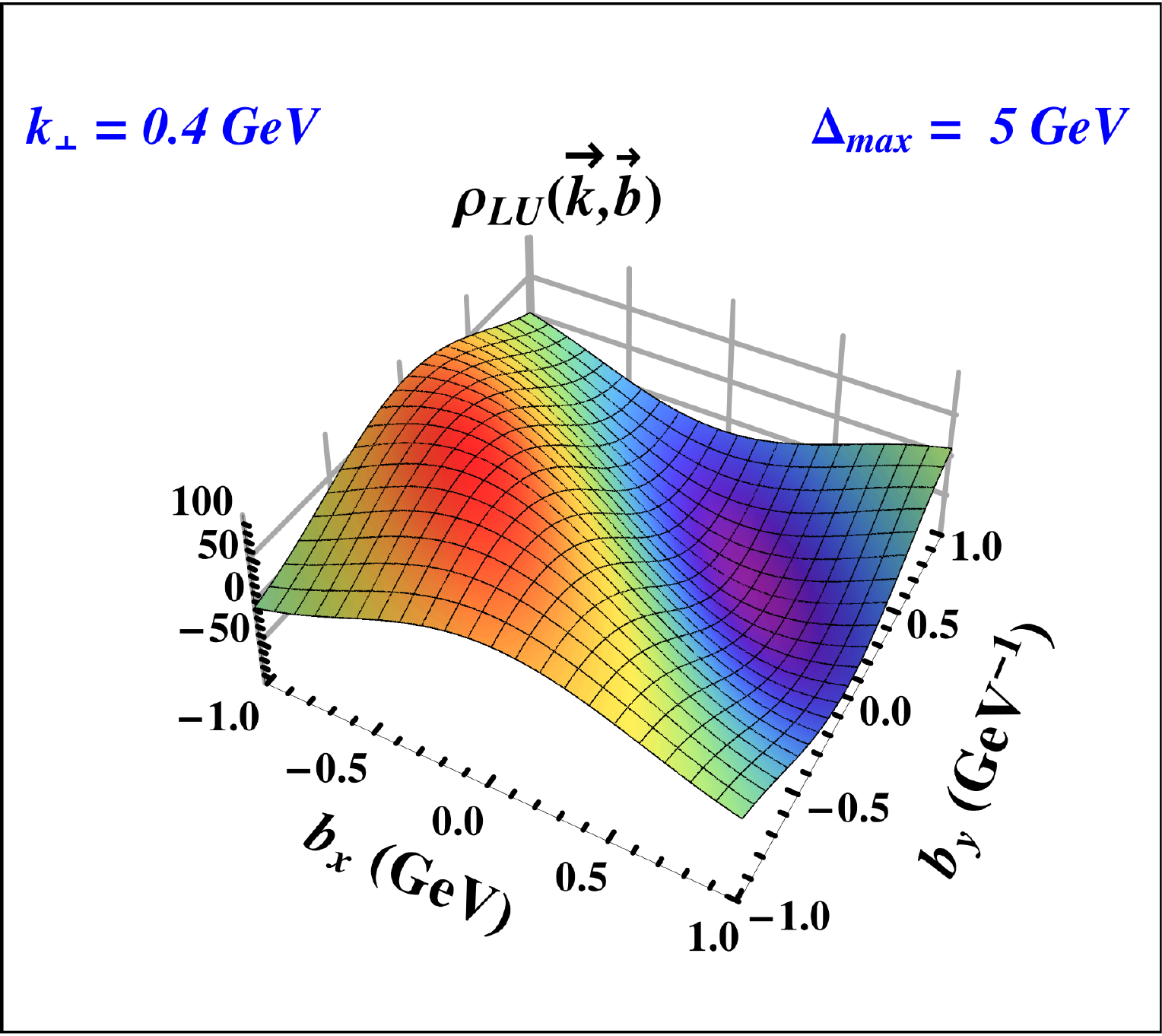}
\caption{\label{fig4}(Color online)
3D plots of the Wigner distributions $\rho_{LU}$ in
$b$ space for $\Delta_\perp^{max} = 5.0$ GeV with $k_\perp = 0.4$ GeV.}
\end{figure}

\noindent
Using the two-particle LFWF the above distributions can be calculated
analytically for a quark state dressed with a gluon \cite{us}. 
In Figs. \ref{fig1} and \ref{fig2} we have shown the 3D plots for the Wigner distribution
$\rho_{UU}$. In
the numerical calculation for Eq.\ref{ruu} we have upper cut-off's
$\Delta_x^{max}$ and
$\Delta_y^{max}$ for the $\Delta_{\perp}$ integration.  In all plots we have
taken $m=0.33$ GeV and integrated over $x$.
We have plotted $\rho_{UU}$ in $b$ space with
$k_\perp = 0.4$ GeV such that  $\vec{k_\perp} = k \hat{j}$. 
The plot has a peak centered at $b_x=b_y=0$ decreasing in the
outer regions of the $b$ space. 
The asymmetry in $b$ space  can be understood from semi-classical arguments in a model with
confinement. As no confining potential is present in our perturbative model, the behavior
is expected to be different. In Figs. \ref{fig3} and \ref{fig4} we show the 3D plots for the
Wigner distribution $\rho_{LU}$.
This is the distortion of the Wigner distribution of unpolarized quarks due
to the longitudinal polarization of the nucleon. We
have shown $\rho_{LU}$ in $b$ space with $k_\perp = 0.4$ GeV such that  
$\vec{k_\perp} = k \hat{j}$ for $\Delta_\perp^{max} = 1.0$ GeV and
$\Delta_\perp^{max} = 5.0$ GeV respectively. Similar to \cite{lorce} we observe a dipole structure in 
these plots and the
dipole magnitude increases with  increase in $\Delta_{max} $.

\section{Orbital Angular Momentum of the quarks}

\noindent
The quark-quark correlator in
Eq.(\ref{qqc}) defining the Wigner distributions  
can be parameterized in terms of generalized transverse momentum dependent
parton distributions (GTMDs) \cite{metz}.
For the twist two case we have eight GTMDs as defined below :

\be
W_{\lambda,\lambda'}^{[\gamma^+]} =
\frac{1}{2M} \bar{u}(p',\lambda') \Bigg[
F_{1,1}-
\frac{i \sigma^{i+}k_{i\perp}}{P^+} F_{1,2}\nn\\ -
 \frac{i\sigma^{i+}\Delta_{i \perp}}{P^+} F_{1,3} +
\frac{i\sigma^{ij}k_{i\perp}\Delta_{j\perp}}{M^2} F_{1,4}
\Bigg] u(p,\lambda);
\label{f11}\ee \\
\be
W_{\lambda,\lambda'}^{[\gamma^+\gamma_5]} =
\frac{\bar{u}(p',\lambda') }{2M} \Bigg[
\frac{-i \epsilon^{ij}_{\perp} k_{i\perp}\Delta_{j\perp}}{M^2}G_{1,1}\nn\\-
\frac{i \sigma^{i+}\gamma_{5} k_{i\perp}}{P^+}G_{1,2} -
 \frac{i\sigma^{i+}\gamma_{5} \Delta_{i\perp}}{P^+} G_{1,3} \nn\\+
i\sigma^{+-} \gamma_5 G_{1,4}
\Bigg] u(p,\lambda).
\label{g11}\ee \\%
\noindent
The GTMDs can be calculated analytically in the dressed quark model. Using
the relation between the GTMD $F_{14}$ and the canonical OAM 
\cite{lorce, hatta1,lorce2}:

\be
l^{q}_z = -\int dx d^{2}k_{\perp} \frac{k_{\perp}^2}{m^2} F_{14}.
\ee

\noindent
The canonical OAM in this model  is given by \cite{us}:
 \be \label{sl}
 l^{q}_{z} = -  2N \int dx (1-x^2) \Big[ I_{1} -  \nn \\m^2(x-1)^2 I_{2}\Big ]
 \ee%

\noindent
On the other hand, the  kinetic quark OAM  is given in terms of the GPDs as
: 
\be 
L^{q}_{z} = \frac{1}{2} \int dx \Bigg\{ 
x  \Big[ H^{q}(x,0,0) + E^{q}(x,0,0)  \Big] \nn\\- \tilde{H^q}(x,0,0)
\Bigg\}. \nn
\ee
In the model considered here, this becomes,
\be \label{cl}
 L^{q}_{z} = \frac{N}{2} \int dx \Big \{
 - f(x) I_{1}  \nn\\+
4m^2(1-x)^2  I_{2}
\Big
\};
\ee
 where,
\begin{align}
I_{1} &=   \int
 \frac{d^{2} k_{\perp}}{m^2(1-x)^2 +(k_{\perp})^2} \nn \\&= \pi
log\Bigg[\frac{Q^2+m^2(1-x)^2}
{\mu^2+m^2(1-x)^2}\Bigg];\nn \\
  I_{2} &=  \int  \frac{d^{2} k_{\perp}}{\Big(m^2(1-x)^2
+(k_{\perp})^2\Big)^2} \nn\\&=
  \frac{\pi}{(m^2(1-x)^2)};\nn\\
  f(x) &=2(1+x^2).
 \nn
 \end{align}

\noindent
$Q$ is the large scale involved in the process, which comes
from the large momentum cutoff in this approach \cite{hari}. 
$\mu$ can be safely taken to be zero provided the quark mass is non-zero. 
Similar qualitative behavior of $L^{q}_z$ and $l^{q}_z$ are observed, both
giving negative values.  
However the magnitude of the two OAM differs in our model, unlike the case
in \cite{lorce}, where these were  calculated in several models
without any gluonic degrees of freedom and  the total
quark contribution to the OAM were equal for both cases. Thus one can see
the effect of the gluonic degrees of freedom to
the OAM in the model considered here.  
The contribution from the gluons has been calculated recently
 also in \cite{other} in this model.           

\section{Conclusion}

\noindent
We presented a recent calculation of  the Wigner distributions for a quark state
dressed with a gluon using the overlap representation in terms of the LFWFs.
This is a relativistic composite spin-1/2 system which has a gluonic degree of
freedom. The Wigner distributions are calculated both for unpolarized and   
longitudinally polarized target and quarks and the correlations are shown in 
transverse position space. The kinetic quark OAM using the
GPD sum rule and the canonical OAM were also calculated in this model 
and it was shown that these are different in  
magnitude, the difference is an effect of the gluonic degree of freedom.   

\begin{acknowledgement}
This work is supported by the  DST project SR/S2/HEP-029/2010, Govt. of
India. AM thanks the organizers of Transversity 2014, June 9-13, 2014, Chia,
Sardinia for the invitation.

\end{acknowledgement}

\end{document}